\documentclass[prb,twocolumn,aps]{revtex4}
\usepackage{graphicx}
\usepackage{dcolumn}
\usepackage{bm}
\usepackage{amsmath,amssymb,amsfonts}
\usepackage{hyperref}
\renewcommand{\vec}[1]{{\mathbf{#1}}}
\newcommand{\beq}{\begin{eqnarray}}
\newcommand{\eeq}{\end{eqnarray}}
\newcommand{\da}{\downarrow}
\newcommand{\ua}{\uparrow}

\begin{document}
\title{Exact Integration of the High Energy Scale in Doped Mott Insulators}
\author{Ting-Pong Choy}
\author{Robert G. Leigh}
\author{Philip Phillips}
\author{Philip D. Powell}
\affiliation{Department of Physics,
University of Illinois
1110 W. Green Street, Urbana, IL 61801, U.S.A.}
\date{\today}

\begin{abstract}
We expand on our earlier work (cond-mat/0612130, Phys. Rev. Lett. {\bf 99}, 46404 (2007) ) in which we constructed the exact low-energy theory of a doped Mott insulator by explicitly integrating (rather than projecting) out the degrees of freedom far away from the chemical potential.  The exact low-energy theory contains degrees of freedom that cannot be obtained from projective schemes. In particular a new charge $\pm 2e$ bosonic field emerges at low energies that is not made out of elemental excitations. Such a field accounts for dynamical spectral weight transfer across the Mott gap. At half-filling, we show that two such excitations emerge which play a crucial role in preserving the Luttinger surface along which the single-particle Green function vanishes.  In addition, the interactions with the bosonic fields defeat the artificial local SU(2) symmetry that is present in the Heisenberg model. We also apply this method to the Anderson-U impurity and show that in addition to the Kondo interaction, bosonic degrees of freedom appear as well.  Finally, we show that as a result of the bosonic degree of freedom, the electron at low energies is in a linear superposition of two excitations--one arising from the standard projection into the low-energy sector and the other from the binding of a hole and the boson.
\end{abstract}

\maketitle

\section{Introduction}
Low energy theories based on an explicit integration over the degrees of
freedom at high energy are the cornerstone\cite{wilson} for analyzing
long wavelength physics of interacting systems.  For high-temperature
superconductivity in the cuprates, the relevant\cite{anderson}
low-energy theory must be constructed for a doped Mott insulator. While
no shortage of theories has been
proposed\cite{Sl1,Sl2,Sl3,Sl4,Sl5,Sl6,Sl7,Sl8,Sl9,Pert0,Pert1,Pert2,
Pert3,Pert33,Pert4,Pert5,Slave1,Slave2,Slave3}, none is based on an
explicit integration over the degrees of freedom at high energy.  The
primary difficulties in carrying out such a program appear in the
simplest model
\beq
H_{\rm Hubb}&=&-t\sum_{i,j,\sigma} g_{ij} c^\dagger_{i,\sigma}c_{j,\sigma}+U\sum_i c^\dagger_{i,\uparrow}c^\dagger_{i,\downarrow}c_{i,\downarrow}c_{i,\uparrow}\nonumber\\&=&H_t+H_U
\eeq
applicable to a doped Mott insulator.  Here $i,j$ label lattice sites,
$g_{ij}$ is equal to one iff $i,j$ are nearest neighbours, $c_{i,\sigma}$
annihilates an electron with spin $\sigma$ on lattice site $i$, $t$ is
the nearest-neighbour hopping matrix element and $U$ the energy cost
when two electrons doubly occupy the same site. The cuprates live in the
strongly coupled regime in which the interactions dominate as $t\approx
0.5$eV and $U\approx 4$eV. As $U$ is the largest energy scale, it is
appropriate to integrate over the fields that generate the physics on
the $U$ scale.  The operators that correspond to such physics can be
written in terms of $\eta_{i,\sigma}=c_{i,\sigma}n_{i,-\sigma}$,  noting
that
$n_{i,\uparrow}n_{i,\downarrow}=\sum_\sigma\eta_{i,\sigma}^\dagger\eta_{i,\sigma}/2$. Physically, $\eta_{i,\sigma}$ ($\eta^\dagger_{i,\sigma}$)
annihilates (creates) an electron on a doubly (singly) occupied site,
hence is associated with the energy scale $U$. Consequently, the
interaction term reduces to a simple quadratic form
\beq
H_U=U\sum_i n_{i,\uparrow}n_{i,\downarrow}=\frac{U}{2}\sum_{i,\sigma}\eta^\dagger_{i,\sigma}
\eta_{i,\sigma},
\eeq
which would enable an exact integration over the $U$ scale {\it if}
$\eta_{i,\sigma}$ obeyed canonical fermionic or bosonic commutation
relations. However, a simple computation gives
\beq
\left\{\eta_i,\eta^\dagger_j\right\}=-\delta_{ij}\frac12\sigma^\mu n_\mu=\left(\begin{array}{c} n_{i,\downarrow}\quad -c_{i,\downarrow}^\dagger c_{i,\uparrow}\\-c_{i,\uparrow}^\dagger c_{i,\downarrow}\quad  n_{i,\uparrow}\end{array}\right).
\eeq
Hence, standard bosonic or fermionic coherent state representations are
of no use in integrating over the fields $\eta_{i,\sigma}$.

The additional problem is spectral weight transfer. When one electron
resides on each site (half-filling), a charge gap of order $U$ opens for
all $U$ in $d=1$ and provided $U\gg t$ for
$d>1$.\cite{anderson,moukouri}   The band above the gap describes
electron motion on singly occupied sites, whereas the band below
captures electron motion on empty sites. Such motion is described by
$\eta_{i,\sigma}$ and $\xi_{i,\sigma}=c_{i,\sigma}(1-n_{i,-\sigma})$,
respectively. However, unlike the traditional band picture in which
electron motion occurs in either the conduction or valence bands,
electron spectral weight lives both above and below the Mott gap.   This
state of affairs obtains because the electron annihilation operator
\beq
c_{i,\sigma}=c_{i,\sigma}(1-n_{i,-\sigma})+c_{i,\sigma}n_{i,-\sigma}=\xi_{i,\sigma}+\eta_{i,\sigma}
\eeq
can be written as a linear combination of excitations that reside in
both bands. That is, unlike the standard band insulator picture, the
states lying above the gap are not orthogonal to those below it.  As a
consequence, adding or removing electrons from a Mott insulator changes
the distribution of spectral weight at all energies.  In particular, the
addition of $x$ holes to a Mott insulator creates at least $2x$ single
particle addition states\cite{sawatzky} just above the chemical
potential.  The deviation from $x$, as would be the case in a band
insulator, is intrinsic to the strong correlations that mediate the Mott
insulating state in a half-filled band, thereby distinguishing Mottness
from ordering. Each hole reduces the number of ways of creating a doubly
occupied site by one, thereby reducing the spectral weight at high
energy.  As the $x$ empty sites can be occupied by either spin up or
spin down electrons, the $2x$ sum rule is exact\cite{sawatzky} in the
atomic limit, $U\to\infty$.  In the presence of hybridization (with
matrix element $t$), virtual excitations between the LHB and UHB
increase the loss of spectral weight at high energy thereby leading to a
faster than $2x$ growth\cite{sawatzky,harris,eskes} of the low-energy
spectral weight, a phenomenon confirmed\cite{cooper,uchida1,chenbatlogg}
widely in the high-temperature copper-oxide superconductors.

A true low-energy is not exact if it cannot account for {\bf all}
low-energy degrees of freedom even if they arise from the high-energy
scale.  Hence, the true low-energy theory of a doped Mott insulator must
preserve the $2x$ sum rule.  In this regard, two approaches are
possible: C1) change the particle statistics so that placing a particle
on one site excludes particles of opposite spin or C2) generate new
degrees of freedom at low energy so that removal of an electron destroys
at least charge $e$ states.  Perturbative methods followed by
projection\cite{Pert0,Pert1,Pert2,Pert3,Pert4} of the high-energy scale
as well as slave\cite{Slave1,Slave2,Slave3} particle techniques all
implement C1. To leading order in $t^2/U$, the result is the $t-J$
model.  The key goal in such approaches is to diagonalize the
Hubbard model into sectors with a fixed number of doubly occupied sites.
When one performs such a transformation\cite{eskes}, however, the electron operators must be transformed as well. Although this step is generally ignored\cite{Sl1,Sl2,Sl3,Sl4,Sl5,Sl6,Sl7,Sl8,Sl9,Pert0,Pert1,Pert2,
Pert3,Pert33,Pert4,Pert5,Slave1,Slave2,Slave3}, it is crucial because the no double occupancy condition applies only to the transformed fermions not to the bare electrons.  As the relationship between the transformed and bare electrons is non-linear, it is advantageous to devise a much simpler method in which the mixing to the doubly occupied sectors in the bare electrons is carried by a single degree of freedom.  The current method provides a solution to this problem.
All the physics associated with the mixing between the UV and IR scales is captured by a charge 2e bosonic field.  In addition, one might entertain the
possibility that slaved-particle methods\cite{Slave1,Slave2,Slave3}
could be tailored to implement an integration of the high-energy scale.
However, in the slaved operator approach, the interactions involving
double occupancy are highly non-linear as a result of the constraints
that remove the unphysical states and hence double occupancy cannot be
integrated over explicitly.

We present here a detailed description of a method that permits an
explicit integration of the degrees of freedom far away from the
chemical potential in doped Mott insulators.   We show that the degrees
of freedom far away from the chemical potential can be integrated out
explicitly (that is, without resorting to projection or slave-particles)
for a doped Mott insulator.  The result is that new physics emerges at
low energies, namely a charge $2e$ boson, that cannot be thought of as
simply related to electronic motion. Our work lays plain that the
true-low energy theory of the Hubbard model is not a $t-J$-like model
in terms of the original bare electrons as is
commonly believed. The true low-energy theory is an example of C2.

The charge $2e$ boson enters the theory as a Lagrange multiplier field.
As such,  it does not have dynamics, in much the same way that the
$\sigma$ field in non-linear $\sigma$-models does not have dynamics. In
that case, dynamics are generated radiatively, by taking into account
interactions with the other fields in the model. In the simplest
spherical case, the latter fields can be completely integrated out, and
a large $N$ expansion organizes the theory. The situation for Mottness
is considerably more complicated. In particular, we have not yet
elucidated the precise low energy dynamics. Instead, we take the
appearance of the $2e$ boson as an indication that the building blocks
for the low energy dynamics of strongly correlated electron matter
involve degrees of freedom that lack electron quantum numbers. As we
show in the present paper, and more fully in a companion
papers\cite{letter,ftmexp}, there are indications that the boson should
not be thought of as a weakly interacting dynamical field at low
energies with, for example, a Fock space of its own, but that instead it
should be thought of as a constituent in a strongly coupled theory.  For
example, from the exact form of the electron creation operator at low
energy, we deduce that the boson can mediate new charge $e$ excitations
by binding a hole.  It is the emergence of this state at low energies
that serves to preserve the $2x$ sum rule\cite{sawatzky}.  We believe
that there are analogies here  between the presence of such composite
states and confining dynamics in particle physics. Indeed, the nature of
an insulator is of course that electric transport is absent, in analogy
to the absence of color transport in QCD.

This work expands considerably our previous Letter\cite{letter} in which
we presented only an outline of the method.

\section{Low-Energy Theory}
We will be concerned with the limit in which the Hubbard bands are
well-separated,  $U\gg t$.  Given that the chemical potential lies in
the gap between such well separated bands at half-filling, which band we
should associate with high energy is ambiguous at half-filling. Both
double occupancy (UHB) and double holes (LHB) are equally costly. 
Doping removes this ambiguity. Hole-doping jumps the chemical potential
to the top of the LHB thereby defining double occupancy to be the high
energy scale. For electron doping, the chemical potential lies at the
bottom of the upper Hubbard band and it is the physics associated with
double holes in the lower Hubbard band that must be coarse-grained.  At
half-filling, both the UHB and LHB must be integrated out.  As each of
these limits results in a different theory, we will present each
separately.  As will be seen, the low-energy theories that result from
the electron and hole-doped cases are related, though not by the naive
particle-hole transformation.

\subsection{Hole Doping}

Within the Hilbert space for the Hubbard model, $\otimes_i \left({\cal
F}_\uparrow\otimes {\cal F}_\downarrow\right)$, it is impossible to
integrate out the degrees of freedom far away from the chemical
potential. The basic idea of our construction is to rewrite the Hubbard
model in such a way as to isolate the high energy degrees of freedom so
that they can be simply integrated out. To solve this problem, it is
expedient to extend the Hilbert space $\otimes_i \left({\cal
F}_\uparrow\otimes {\cal F}_\downarrow\otimes {\cal F}_D\right)$. The
key idea is to associate $D^\dagger$ with the creation of
double-occupation, to be implemented by a constraint.  In order to limit
the Hilbert space to single occupation in the $D$ sector, we will take
$D$ to be fermionic.  We refer to $D$ as a fermionic oscillator as it is
associated to a two state system.  The field $D$ will enter the theory
as an elemental field with a large (order $U$) quadratic term and
precise interactions with the electronic degrees of freedom; the
low-energy (IR) theory is obtained by integrating out $D$. The
interactions of this extended model must be chosen so that the model is
precisely equivalent to the Hubbard model; indeed, if instead of
integrating out the field $D$, we merely solve the aforementioned
constraint, the model will reduce to the Hubbard model, which we will
refer to as the high energy (UV) theory.

\begin{figure}
\centering
\includegraphics[width=10.8cm]{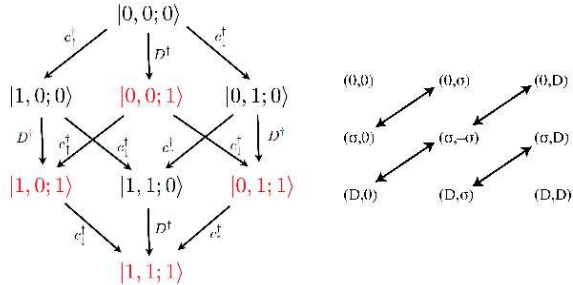}
\caption{(a) extended Hilbert space for a single site.
 (b)  hopping processes between neighbouring sites included in the Lagrangian. Double occupation has been replaced by $D$-occupation.}
\label{hilb}
\end{figure}

The action of the standard electron creation operator,
$c^\dagger_{i,\sigma}$ and the new fermionic operator, $D^\dagger$, to
create the allowed states on a single site are shown in Fig. \ref{hilb}.

There are of course several unphysical states in this Hilbert space. As
we will see, such states are removed once the constraint is solved.  At
present, the expansion of the Hilbert space should be thought of as a
tool to enable the integration of the high energy degrees of freedom. To
proceed, we formulate a Lagrangian in the extended Hilbert space. The
allowed hops involving the $D$ fields and the electron operators which
are equivalent to the hops in the Hubbard model are indicated in
Fig(\ref{hilb}b). For example, the hopping process in the upper
left-hand corner describes the hopping of a hole in the lower-Hubbard
band.  The terms in the middle describe the transport between the $D$
field and two electrons in a singlet on neighbouring sites.  The term in
the lower right corner describes a hop in which $D_i$ and
$c_{j}^\dagger$ switch places. There are no further allowable hopping
processes.  A further requirement of the Lagrangian for the hole-doped
theory is that it contain the appropriate dynamical term for motion in
the lower Hubbard band. That is,  those sites which contain the
occupancy $c^\dagger_{i,\downarrow}c^\dagger_{i,\uparrow}|0\rangle$ must
be excluded from hopping processes (such hops are accounted for by the
hopping of $D$).  The Euclidean Lagrangian in the extended Hilbert space
which describes the hopping processes detailed above can be written
\beq\label{LE}
L&&=\int d^2\theta\left[\bar{\theta}\theta\sum_{i,\sigma}(1- n_{i,-\sigma}) c^\dagger_{i,\sigma}\dot c_{i,\sigma} +\sum_i D_i^\dagger\dot D_i\right.\nonumber\\
&&+U\sum_j D^\dagger_jD_j- t\sum_{i,j,\sigma}g_{ij}\left[ C_{ij,\sigma}c^\dagger_{i,\sigma}c_{j,\sigma}
+D_i^\dagger c^\dagger_{j,\sigma}c_{i,\sigma}D_j\right.\nonumber\\
&&+\left.\left.(D_j^\dagger \theta c_{i,\sigma}V_\sigma c_{j,-\sigma}+h.c.)\right]+H_{\rm con}\right].
\eeq
Here, $g_{ij}$ selects out nearest neighbours (note that if we wanted to include next-to-nearest neighbour interactions, we need just modify the matrix $g_{ij}$ accordingly), the parameter $V_\sigma$
has values $V_\uparrow =1$, $V_\downarrow=-1$, and simply ensures that
$D$ couples to the spin singlet and the operator $C_{ij,\sigma}$ is of
the form
$C_{ij,\sigma}\equiv\bar\theta\theta\alpha_{ij,\sigma}\equiv\bar\theta\theta(1-n_{i,-\sigma})(1-n_{j,-\sigma})$ with number operators
$n_{i,\sigma}\equiv c^\dagger_{i,\sigma}c_{i,\sigma}$.

For simplicity, we have introduced a complex Grassman constant $\theta$,
which we have inserted in order to keep track of statistics; it bears
some resemblance to a superspace coordinate. Because $D_j$ is fermionic
and $c_{j,\uparrow}c_{j,\downarrow}$ transforms as a boson, a Grassman
variable is needed to essentially `fermionize' double occupancy. They
are normalized via
\beq
\int d^2\theta\ \bar{\theta}\theta=1.
\eeq
The Grassmann variable is an artificial device that will disappear from
the UV or IR Lagrangians.

The constraint Hamiltonian $H_{\rm con}$ is taken to be
\beq\label{con}
H_{\rm con} = s\bar{\theta}\sum_j\varphi_j^\dagger (D_j-\theta c_{j,\uparrow}c_{j,\downarrow})+h.c.
\eeq
where $\varphi$ is a complex charge $2e$ bosonic field which enters the
theory as a Lagrange multiplier.  The constant $s$ has been inserted to
carry the units of energy.
At this point, there is some ambiguity in the normalization of
$\varphi$, but we expect that this will be set dynamically. We will find
that if a true infrared limit exists, then $s$ must be of order the
hopping matrix element $t$.  There is a natural parallel between the
constraint condition, Eq. (\ref{con}), and the constraint in the
non-linear sigma model.   In fact, the auxiliary field $\varphi$ will
enter the low-energy theory in an analogous fashion to $\sigma$ in the
non-linear sigma model. In both cases, $\varphi$ and $\sigma$ enter as
Lagrange multipliers.  Both end up playing a crucial role in the phase
structure of the true low-energy or infrared limit.  In this case,
$\varphi$ will serve to create new excitations at low energy which will
generate the dynamical part of the spectral weight transfer across the Mott gap.

Now, as remarked previously, we have chosen the Lagrangian (\ref{LE}) so
that this theory is equivalent to the Hubbard model. To demonstrate
this, we first show that once the constraint is solved, we obtain the
Hubbard model. Hence, the Lagrangian we have formulated is the Hubbard
model written in a non-traditional form -- in some sense, we have
inserted unity into the Hubbard model path integral in a rather
complicated fashion.  To this end, we compute the partition function
\beq\label{Z}
Z=\int [{\cal D}c\ {\cal D}c^\dagger\ {\cal D}D\ {\cal D}D^\dagger\ 
{\cal D}\varphi\ {\cal D}\varphi^\dagger]\exp^{-\int_0^\tau L dt}.
\eeq
with $L$ given by (\ref{LE}). We note that $\varphi$ is a Lagrange
multiplier. As shown in the Appendix (Eq. (\ref{lm})), in the Euclidean
signature, the fluctuations of the real and imaginary parts of
$\varphi_i$ must be integrated along the imaginary axis for $\varphi_i$
to be regarded as a Lagrangian multiplier. The $\varphi$ integrations
(over the real and imaginary parts) are precisely a representation of (a
series of) $\delta$-functions of the form,
\beq
\delta\left(\int d\theta D_i-\int d\theta\ \theta c_{i,\uparrow}c_{i,\downarrow}\right).
\eeq

If we wish to recover the Hubbard model, we need only to integrate over
$D_i$, which is straightforward because of the $\delta$-functions.  The
dynamical terms yield
\beq
&&\int d^2\theta\ \bar\theta\theta\left[\sum_{i,\sigma}(1- n_{i,-\sigma}) 
c^\dagger_{i,\sigma}\dot c_{i,\sigma}+\sum_i c^\dagger_{i,\downarrow}
c_{i,\uparrow}^\dagger\dot c_{i,\uparrow}c_{i,\downarrow}\right.\nonumber\\
&&\qquad\qquad\left.+\sum_ic^\dagger_{i,\downarrow}c_{i,\uparrow}^\dagger 
c_{i,\uparrow}\dot c_{i,\downarrow}\right]\nonumber\\
&&=\int d^2\theta\ \bar\theta\theta\sum_{i,\sigma}\left[(1-n_{i,-\sigma})
c^\dagger_{i,\sigma}\dot c_{i,\sigma}+n_{i,-\sigma}c_{i,\sigma}^\dagger
\dot c_{i,\sigma}\right]\nonumber\\
&&=\int d^2\theta\ \bar\theta\theta\sum_{i,\sigma}c_{i,\sigma}^\dagger \dot c_{i,\sigma}.
\eeq
Likewise the term proportional to $V_\sigma$ yields
\beq\label{den1}
&&\int d^2\theta\ \bar\theta\theta \sum_{i,j}g_{ij}\left[c_{j,\downarrow}^\dagger 
c_{j,\uparrow}^\dagger(c_{i,\uparrow}c_{j,\downarrow}-c_{i,\downarrow}c_{j,\uparrow})\right]
+h.c.\nonumber\\
&&=\int d^2\theta\ \bar\theta\theta \sum_{i,j,\sigma}g_{ij}n_{j,-\sigma}
c^\dagger_{j,\sigma}c_{i,\sigma}+h.c.
\eeq
Finally, the hopping terms that involve two $D$ fields give rise to
\beq\label{den2}
&&\int d^2\theta\ \bar\theta\theta\sum_{i,j}g_{ij}\left[c_{i,\downarrow}^\dagger c_{i,\uparrow}^\dagger(c_{j,\uparrow}^\dagger c_{i,\uparrow}+c_{j,\downarrow}^\dagger c_{i,\downarrow})c_{j,\uparrow}c_{j,\downarrow}\right]\nonumber\\
&&=-\int d^2\theta\ \bar\theta\theta\sum_{i,j}g_{ij}n_{j,-\sigma}n_{i,-\sigma}c_{i,\sigma}^\dagger c_{j,\sigma}.
\eeq
Eqs. (\ref{den1}) and (\ref{den2}) add to the constrained hopping term
in the Lagrangian (the term proportional to $C_{ij,\sigma}$) to yield the
standard kinetic energy term in the Hubbard model. Finally, the
$D^\dagger D$ term  generates the on-site repulsion of the Hubbard
model.  Consequently, by integrating over $\varphi_i$ followed by an
integration over $D_i$, we recover the Lagrangian,
\beq
\int d^2\theta\ \bar\theta\theta L_{\rm Hubb}=\sum_{i,\sigma}c_{i,\sigma}^\dagger\dot c_{i,\sigma}+H_{\rm Hubb},
\eeq
of the Hubbard model.  This constitutes the ultra-violet (UV) limit of
our theory. In this limit, it is clear that the Grassman variables
amount to an insertion of unity and hence play no role.  Further, in
this limit the extended Hilbert space contracts, unphysical states such
as  $|1,0,1\rangle$, $|0,1,1\rangle$, $|1,1,1\rangle$ are set to zero,
and we identify $|1,1,0\rangle$ with $|0,0,1\rangle$.  Note there is no
contradiction between treating $D$ as fermionic and the constraint in
Eq. (\ref{con}). The constraint never governs the commutation relation
for $D$. The value of $D$ is determined by Eq. (\ref{con}) only when
$\varphi$ is integrated over. This is followed immediately by an
integration over $D$ at which point $D$ is eliminated from the theory.

The advantage of our starting Lagrangian over the traditional writing of
the Hubbard model is that we are able to coarse grain the system cleanly
for $U\gg t$.  The energy scale associated with $D$ is the large on-site
energy $U$.  Hence, it makes sense, instead of solving the constraint,
to integrate out $D$.  The resultant theory will contain explicitly the
bosonic field, $\varphi$.   As  a result of this field, double occupancy
will remain, though the energy cost will be shifted from the UV to the
infrared (IR).  Because the theory is Gaussian, the integration over
$D_i$ can be done exactly.  This is the ultimate utility of the
expansion of the Hilbert space -- we have isolated the high energy
physics into this Gaussian field. As a result of the dynamical term in
the action, integration over $D$ will yield a theory that is frequency
dependent. The frequency will enter in the combination $\omega+U$ which
will appear in  denominators.  Since $U$ is the largest energy scale, we
expand in powers of $\omega/U$; the leading term yields the proper
$\omega=0$ low-energy theory.   Since the theory is Gaussian, it
suffices to complete the square in the $D$-field. To accomplish this, we
define the matrix
\beq\label{eom}
{\cal M}_{ij}=\delta_{ij}-\frac{t}{(\omega+U)}g_{ij}
\sum_\sigma c_{j,\sigma}^\dagger c_{i,\sigma}
\eeq
and $b_{i}=\sum_{j}b_{ij}=\sum_{j,\sigma} g_{ij}c_{j,\sigma}V_\sigma
c_{i,-\sigma}$. At zero frequency the Hamiltonian is
\beq
H^{IR}_h = -t\sum_{i,j,\sigma}g_{ij} \alpha_{ij,\sigma}
c^\dagger_{i,\sigma}c_{j,\sigma}+ H_{\rm int}-\frac{1}{\beta}Tr\ln{\cal M},
\nonumber
\eeq
where
\beq\label{HIR}
H_{\rm int}=-\frac{t^2}U \sum_{j,k} b^\dagger_{j} ({\cal M}^{-1})_{jk} b_{k}
-\frac{s^2}U\sum_{i,j}\varphi_i^\dagger
 ({\cal M}^{-1})_{ij} \varphi_j\nonumber\\
-s\sum_j\varphi_j^\dagger c_{j,\uparrow}c_{j,\downarrow}
+\frac{st}U \sum_{i,j}\varphi^\dagger_i ({\cal M}^{-1})_{ij}
b_{j}+h.c.
\eeq
which constitutes the true (IR) limit as long as the energy scale $s$ is
not of order $U$. If $s\sim O(U)$ then we should also integrate out
$\varphi_i$ -- this integration is again Gaussian and can be done
exactly; one can easily check that this leads precisely back to the UV
theory, the Hubbard model.\footnote{Thus, the order of integration does
not matter, as one would expect.}  Hence, the only way in which a
low-energy theory of the Hubbard model exists is if the energy scale for
the dynamics that $\varphi$ mediates is $O(t)$.   This observation is
significant because it lays plain the principal condition for the
existence of an IR limit of the Hubbard model.

To fix $s$, we note that the fourth term entering our Hamiltonian can
mediate spin exchange. As the energy scale for this process is $t^2/U$,
we make the identification $s\simeq t$.  Hence, the low-energy theory
contains a charge $2e$ bosonic field which can either annihilate/create
doubly occupied sites or nearest-neighbour singlets.  That the energy
cost for double occupancy in the IR is $t^2/U$ and not $U$ underscores
the fact that the UHB and LHB are not orthogonal.  The presence of the
new field $\varphi_i$ at low energies is the result of the overlap
between the high and low energy scales. Physically, double occupancy
occurs at low energies for two distinct reasons.  The first is spin
exchange which is generated by the term $\varphi_i b_i^\dagger$.  The
second is motion of a doubly occupied site (a doublon) entirely in the
LHB.  The latter is not present in projective models but is certainly a
low-energy process that must be present in the exact low-energy theory.

While electron number conservation is broken in the IR, we find (by
inspection of (\ref{HIR})) that a conserved low-energy charge does
exist\cite{charge}
\beq
Q=\sum_{i,\sigma}c_{i,\sigma}^\dagger c_{i,\sigma}+2\sum_i\varphi^\dagger_i\varphi_i.
\eeq
As Eq. (\ref{HIR}) makes clear, bosons acquire dynamics only through
electron motion.  Hence, the low-energy theory of a hole-doped Mott
insulator is a strongly coupled bose-fermi model. On purely
phenomenological grounds, bose-fermi models have been
advanced\cite{tdlee,ranninger} as a starting point for tackling the
cuprate problem. In such models, and others\cite{Pert3}, the bosons are
viewed as non-interacting and possess a Fock space of their own. 
However, the current analysis lays plain that while the bosonic degree
of freedom exists, it does not extend the Hilbert space of the Hubbard
model.  That is, the charge $2e$ boson does not have a Fock space of its
own.  In obtaining the low-energy theory, we integrated over the high
energy $D-$ field which acted in the extended Hilbert space. 
Consequently, the resultant low-energy theory preserves the original
Hilbert space of the Hubbard model. As we will show, a distinct
possibility is that the boson acts to create composite excitations that
have charge $e$.

Several limits are of interest.  First, consider the limit  $U=\infty$
(for fixed lattice size).  The theory reduces to the restricted hopping
term and the third term in Eq. (\ref{HIR}).  In this limit, the
$\varphi$ integration reduces to a delta function,
$\delta(c_{i,\uparrow}c_{i,\downarrow})$, giving a constraint enforcing
the vanishing of double occupancy, the correct result for $U=\infty$.

Second, should $\varphi=0$, we recover the interactions in the $t-J$
model.  To establish this, we note that for $\varphi_i=0$, we have the
restricted hopping term and the first term in Eq. (\ref{HIR}). 
Approximating ${\cal M}_{ij}$ by its leading term, $\delta_{ij}$, the
second term reduces to
\beq
\sum_i b_i^\dagger b_i=\sum_{ij\ell\sigma\sigma'} g_{ij}g_{\ell i} 
c^\dagger_{i,-\sigma}V_\sigma
c^\dagger_{j,\sigma} c_{\ell,\sigma'} V_{\sigma'} c_{i,-\sigma'}
\eeq
 which contains the spin-spin interaction $-(\vec S_i\cdot \vec S_j-n_i
 n_j/4)$ as well as the three-site hopping term. Next, we expand the
 $Tr\ln$ term
\beq
Tr\ln{\cal M}^{-1} &=& Tr\ln (\delta_{ij}+  A_{ij})\nonumber\\
&=& A_{ii} + \frac{1}{2} A_{ij} A_{ji}+\ldots\nonumber\\
&=&  \frac{t^2}{(\omega + U)^2} \sum_{\sigma,\sigma'} \sum_{ i,j}g_{ij}
c_{j,\sigma}^\dagger c_{i,\sigma} c_{i,\sigma'}^\dagger c_{j,\sigma'} +\ldots\nonumber
\eeq
where
\beq
A_{ij}= \frac{t}{\omega+U}\sum_\sigma g_{ij}c_{j,\sigma}^\dagger c_{i,\sigma}
\eeq
is nonzero only if $i,j$ are nearest neighbours. When this term appears
in the Euclidean Lagrangian, its magnitude is $k_BT t^2/(\omega+U)^2 $. 
Therefore, at low temperature, this term is small compared to $H_{\rm
IR}$ and to leading order in $t^2/U$, the terms in $H_{\rm IR}$
dominate. Hence, the $\varphi=0$ limit contains the interactions in the
$t-J$ model, thereby establishing that the physics contained in
$\varphi_i$ is non-projective. To make closer contact with the $t-J$
model in which the spin-spin interaction acts only in the
singly-occupied sector, we note that the theory we have developed here
could have been formulated strictly in the projected space by simply
substituting $\xi_{i,\sigma}$ for $c_{i,\sigma}$ in the hopping terms
containing $D_i$ in our starting Lagrangian, Eq. (\ref{LE}). The only
substantive difference would be that the second hopping term (the term
quadratic in $D_i$) in Eq. (\ref{LE}) would enter with the opposite
sign.  Hence, in the IR limit, ${\cal
M}_{ij}\rightarrow\delta_{ij}+t/U\xi_{j,\sigma}^\dagger \xi_{i,\sigma}$. 
This change is dictated by the commutation relations of the
$\xi_{i,\sigma}$ operators. The UV limit, the Hubbard model, is obtained
as before.  Setting $\varphi_i=0$ in the IR limit leads exactly to the
$t-J$ model.  Thus, the $t-J$ model\cite{Pert33} written in terms of the bare electron operators is not the low-energy
limit of the Hubbard model. This is not entirely surprising as Eskes and others\cite{eskes} has stressed that the operators must be transformed as well in writing the $t-J$ model.
Only at $U=\infty$ do the transformed and bare fermion operators agree.
Hence, at any finite $U$, the physics is governed by a finite length scale for
double occupancy. Hence the limits $U\rightarrow\infty$ and
$L\rightarrow\infty$ (L the size of the system) do not commute as is
required for a hard projective model (no double occupancy in the original electron basis) to be the true low-energy theory of the
Hubbard model.

No such problem besets our low-energy theory. We can recover the
original Hubbard model from our low-energy theory by simply integrating
over $\varphi_i$. Although this is not a sensible thing to do from a low
energy perspective, it can be done exactly.  To see how this happens, we
rewrite  Eq. (\ref{HIR})  including the frequency dependence,
\beq
L_{IR} &=&\sum_{i,\sigma}(1-n_{i\bar\sigma})c_{i,\sigma}^\dagger \dot c_{i,\sigma} -t\sum_{i,\sigma} (1-n_{i\bar\sigma})c_{i,\sigma}^\dagger c_{j,\sigma} (1-n_{j\bar\sigma}) \nonumber\\
&-& \sum_{ij}\left(  s\varphi_i^\dagger - t\sum_{\langle i,k\rangle}b_{ki}^{\dagger} +(\omega +\hat U) c^\dagger_{k\downarrow} c^\dagger_{k\uparrow} {\cal M}_{ki} \right)\nonumber\\
&&\times  \frac{({\cal M}^{-1})_{ij}}{\omega + \hat U} \left(  s\varphi_j  - t\sum_{\langle j,n\rangle}b_{nj}+(\omega+ \hat U) {\cal M}_{jn} c_{n\uparrow} c_{n\downarrow}  \right) \nonumber\\
&-& t c_{i,\downarrow}^\dagger c_{i,\uparrow}^\dagger b_i   - tb_i^{\dagger} c_{i,\uparrow} c_{i,\downarrow}  + (\omega+\hat U ) c_{i,\downarrow}^\dagger c_{i,\uparrow}^\dagger {\cal M}_{ij}  c_{j,\uparrow} c_{j,\downarrow}\nonumber\\&-&\frac{1}{\beta}Tr\ln{\cal M}\nonumber
\eeq
To simplify the expression, note that,
\beq
- t b_i^{\dagger} c_{i,\uparrow}c_{i,\downarrow} &=& -t \sum_j (c_{i,\downarrow}^\dagger c_{j,\uparrow}^\dagger - c_{i,\uparrow}^\dagger c_{j,\downarrow}^\dagger) c_{i,\uparrow}c_{i,\downarrow}\nonumber\\
&=& -t \sum_\sigma  c_{j,\sigma}^\dagger c_{i,\sigma} n_{i,-\sigma}
\eeq
and
\beq
-\sum_{\langle i,j\rangle\sigma} c_{i,\downarrow}^\dagger c_{i,\uparrow}^\dagger  c_{j,\sigma}^\dagger c_{i,\sigma} c_{j,\uparrow} c_{j,\downarrow}= \sum_{\langle i,j\rangle,\sigma} n_{i,-\sigma} c_{i,\sigma}^\dagger c_{j,\sigma} n_{j,-\sigma}.
\eeq
which comes from $c_{i,\downarrow}^\dagger c_{i,\uparrow}^\dagger {\cal M}_{ij}  c_{j,\uparrow} c_{j,\downarrow}$. The Lagrangian becomes
\beq
L&=&\sum_{i,\sigma} c_{i,\sigma}^\dagger \dot c_{i,\sigma} + H_{Hubb}\nonumber\\
&-& \sum_{ij}\left(  s\varphi_i^\dagger - t\sum_{\langle i,k\rangle}b_{ki}^{\dagger} +(\omega +\hat U) c^\dagger_{k\downarrow} c^\dagger_{k\uparrow} {\cal M}_{ki} \right)\nonumber\\
&& \frac{({\cal M}^{-1})_{ij}}{\omega + U} \left(  s\varphi_j - t\sum_{\langle j,n\rangle}b_{nj} +(\omega + \hat U) {\cal M}_{jn} c_{n\uparrow} c_{n\downarrow}  \right)\nonumber\\
&&-\frac{1}{\beta}Tr\ln{\cal M}
\eeq
which yields the Hubbard model upon integration over $\varphi_i$ (see
the Appendix for details).

We conclude from this analysis that our low-energy theory permits
immediate correspondence with the original Hubbard model; that is, we
have not lost any information regarding the high energy scale, unlike
projective methods.  All information regarding the high-energy scale is
encoded into the emergent charge $2e$ bosonic excitation and its
interactions. The IR physics will be determined by examining the low
energy dynamics of the electrons/holes and $\varphi$.

\subsection{What this theory is not}

Rather than decoupling the on-site repulsion term, we derived our
low-energy theory by exponentiating a $\delta-$functional constraint on
the heavy field, $D$. Nonetheless, one might contemplate that our theory
could be obtained by more traditional schemes, for example by some sort
of Hubbard Stratonovich (HS) decoupling scheme.  Since the interaction
in the Hubbard model is entirely local, any decoupling by means of
introducing an auxiliary field  would yield only local interactions. 
The auxiliary field $\varphi_i$ in Eq. (\ref{HIR}) clearly generates
non-local interactions as well as on-site interactions.  Hence, Eq.
(\ref{HIR}) cannot be obtained from a HS transformation. However, the
non-local terms are scaled by $t/U$. Hence, it might still be maintained
that the local terms dominate and in fact that they could be obtained by
some sort of HS transformation.   Consider the identity,
\beq\label{iden}
e^{-UX^\dagger X}=\int d^2\sigma\ e^{-(\lambda \sigma^\dagger\sigma + \lambda_1 \sigma X^\dagger + \lambda_2X \sigma^\dagger)}
\eeq
which is true as long as  $\lambda\in \mathbb{R}^+$ and
$\lambda_1\lambda_2=-U\lambda$.  The standard HS transformation assumes
that $\lambda_1\lambda_2\in\mathbb{R}^+$ and $\lambda_2=\lambda_1^\ast$.
 This necessarily leads only to the $-U$ Hubbard model with
$X=c_\uparrow c_\downarrow$.  However, the constraint that the exponent
on the right-hand side of Eq. (\ref{iden}) be real can be
relaxed\cite{complex} in which case it can be applied to the $+U$
Hubbard model as well.  Nonetheless, this procedure will not yield the
non-local terms in our low-energy theory. Further, it does not permit a
clean identification of the field associated with the high-energy
degrees of freedom. That is, there is no D-field in any version of Eq.
(\ref{iden}). Consequently, this procedure is a non-starter for the
construction of a proper low-energy theory.

\subsection{Electron doping}

For electron doping, the chemical potential jumps to the bottom of the
UHB.  Consequently, the degrees of freedom that lie far away from the
chemical potential no longer correspond to double occupancy but rather
double holes. To coarse grain these degrees of freedom, we extend the
Hilbert space in a similar way as the hole-doped theory, defining a new
field $\tilde D$ which will be constrained to describe the creation of
double holes. Mathematically, all results in hole doping obtained in the
previous section can be transformed to electron doping via a generalized
particle-hole transformation (GPHT), namely,
\beq
\label{eq:PH-UV}
c_{i,\sigma}&\rightarrow& e^{i\vec Q  \cdot \vec R_i} c_{i,\sigma}^\dagger\nonumber\\
D_i&\rightarrow& \tilde D_i\nonumber\\
\varphi_i &\rightarrow& \tilde \varphi_i
\eeq
where $\vec Q = (\pi,\pi)$ and $\tilde D_i$ is a fermion operator
associated with double holes. The new bosonic field, $\tilde\varphi$, is
the Lagrangian multiplier defined by the constraint, $\tilde{H}_{\rm
con}=\tilde s\bar{\theta}\sum_i\tilde\varphi_i(\tilde D_i-\theta
c_{i,\uparrow}^\dagger c_{i,\downarrow}^\dagger)+h.c.$. According to Eq.
(\ref{LE}), an appropriate Lagrangian for the extended theory at
electron doped can be constructed,
\beq\label{Led} L&&=\int
d^2\theta\left[\bar{\theta}\theta\sum_{i,\sigma}n_{i,-\sigma}
 c_{i,\sigma}\dot c_{i,\sigma}^\dagger + \sum_i \tilde D_i^\dagger \dot{\tilde D}_i + U\sum_j \tilde D_j^\dagger \tilde D_j\right.\nonumber\\
&&- t\sum_{i,j,\sigma}g_{ij}\left[c^\dagger_{i,\sigma}n_{i,-\sigma}c_{j,\sigma}n_{j,-\sigma} +\tilde D_j^\dagger c^\dagger_{j,\sigma}c_{i,\sigma}\tilde D_i\right.\nonumber\\
&&+\left.\left.(\bar\theta c_{i,\sigma}V_\sigma c_{j,-\sigma}\tilde D_i+h.c.)
\right]+\tilde{H}_{\rm con}\right],
\eeq
that preserves the distinct hops in the Hubbard model. Two differences
to note are that 1) because the chemical potential resides in the UHB,
the electron hopping term now involves sites that are at least singly
occupied and 2) the order of the $\tilde D_i$ and $c_i$ operators is
important.  If we integrate over $\tilde\varphi_i$ and then $\tilde
D_i$, all the unphysical states are removed and we obtain as before
precisely $L_{\rm Hubb}$. Hence, both theories yield the Hubbard model
in their UV limits.  They differ, however, in the IR as can be seen by
performing the integration over $\tilde D_i$.  The corresponding
integral is again Gaussian and yields
\beq\label{edir}
H^{IR}_{\rm e} = -t\sum_{i,j,\sigma}
g_{ij}c^\dagger_{i,\sigma}n_{i,-\sigma}c_{j,\sigma}n_{j,-\sigma}+
\tilde{H}_{\rm int}-\frac{1}{\beta}Tr\ln{\cal \tilde M}\nonumber
\eeq
where
\beq\label{HIRE}
\tilde{H}_{\rm int}&=&-\frac{t^2}U \sum_{j,k}
b_{j}  ({\cal \tilde M}^{-1})_{jk}
b^\dagger_{k}-\frac{\tilde s^2}U\sum_{i,j}\tilde\varphi_i^\dagger
({\cal \tilde M}^{-1})_{ij} \tilde\varphi_j\nonumber\\
&&-\tilde s\sum_j \tilde\varphi_j^\dagger c_{j,\uparrow}^\dagger c_{j,\downarrow}^\dagger + \frac{\tilde s t}U \sum_{i,j} \tilde\varphi^\dagger_i ({\cal \tilde M}^{-1})_{ij}
b_{j}^\dagger + h.c.\nonumber
\eeq
as the IR limit of the electron-doped theory. In Eq.(\ref{HIRE}), the
matrix ${\cal \tilde M}_{ij}={\cal M}_{ji}$ is defined via the GPHT on
the ${\cal M}$ matrix in Eq.(\ref{eom}). As $\tilde\varphi_i$ is a
complex field, the GPHT interchanges the creation operators of opposite
charge. We again make the identification $\tilde s\sim t$ because the
last term can also mediate spin exchange.

When the boson vanishes, we do recover the exact particle-hole symmetric
analogue of the hole-doped theory.  Because the field $\tilde\varphi$
now couples to double holes, the relevant creation operator has charge
$-2e$ and the conserved charge is
$\tilde{Q}=\sum_{i,\sigma}c_{i,\sigma}^\dagger
c_{i,\sigma}-2\varphi_i^\dagger\varphi_i$.  This sign change in the
conserved charge will manifest itself as a sign change in the chemical
potential as long as
$\langle\tilde\varphi_i^\dagger\tilde\varphi_i\rangle\ne 0$. Likewise,
the correct $U\rightarrow\infty$ limit is obtained as before.

\subsection{Half Filled Models}

At half filling when the chemical potential lies within the Mott gap,
both double hole and double occupancy lie far away from the chemical
potential. As two different degree of freedom per site need to be coarse
grained, we introduce two new fields $D$ and $\tilde D$ which when
constrained correspond to double occupancy and double holes,
respectively and extend the Hilbert space to $\otimes_i \left({\cal
F}_\uparrow\otimes {\cal F}_\downarrow\otimes {\cal F}_D \otimes {\cal
F}_{\tilde D}\right)$. The corresponding low-energy theory will be
obtained by integrating over both $D$ and $\tilde{D}$ rather than by
solving the constraint.  As a result, integrating over $D$ and
$\tilde{D}$ do not yield identical results as would be the case if
double occupancy or double holes were integrated over.

\subsubsection{Anderson Impurity}

To illustrate the process of coarse graining two high energy fields, we
begin with a simpler model, the Anderson\cite{aimp} impurity model.
\beq
\label{anderson}
H&=&E_f (n_\uparrow + n_\downarrow) + U n_\downarrow n_\uparrow \nonumber\\&+& \sum_{\vec k,\sigma} n_{\vec k, \sigma} \varepsilon_{\vec k} + \sum_{\vec k} \left(v_{\vec k} c_{\vec k, \sigma}^\dagger a_\sigma + h.c. \right)
\eeq
where $a_\sigma$ destroys an impurity electron and $c_{\vec k,\sigma}$
destroys a continuum electron. By setting $E_f=-U/2$, it costs an energy
$U/2$ to create an double hole or a doubly occupied states on the
impurity site which is analogous to the half-filled Hubbard model. In
the following, we would like to show, by introducing two heavy fields
$D$ and $\tilde D$ which correspond to the doubly occupied or double
hole states on the impurity site respectively, the Kondo model with
additional coupling to new bosonic fields can be derived perturbatively
if $v_{\vec k} \ll U$.

The appropriate extended Hamiltonian is
\begin{widetext}
\beq
H&=&\int d^2\theta  \left[E_f \sum_\sigma n_\sigma + \frac{1}{2}UD^\dagger D -\frac{1}{2}U \tilde D \tilde D^\dagger + \bar\theta\theta \sum_{\vec,\sigma}n_{\vec k,\sigma}\varepsilon_{\vec k}
+\sum_{\vec k,\sigma} \left(\bar\theta c_{\vec k,-\sigma}^\dagger V_\sigma a_\sigma^\dagger D + \bar{D}^\dagger\theta c_{\vec k,-\sigma}^\dagger V_\sigma a_\sigma^\dagger + h.c.  \right)\right.\nonumber\\
&&\left.+s\bar\theta\varphi^\dagger \left(D-\theta a_\uparrow a_\downarrow\right)
 +h.c. +\tilde s \bar\theta \tilde \varphi^\dagger \left(\tilde D-\theta a_\uparrow^\dagger a_\downarrow^\dagger\right) +h.c.  \right],
\eeq
\end{widetext}
In the current model, two bosonic field $\varphi$ and $\tilde\varphi$
are introduced which corresponding to the two constraints on $D$ and
$\tilde D$ fields respectively. If we first integrate out $\varphi$,
$\tilde\varphi$ which result in delta functions corresponding to the
removal of the unphysical states and then integrate out $D$ and $\tilde
D$, Eq. (\ref{anderson}) is obtained precisely. This constitutes the
ultra-violet (UV) limit of this theory. Similar to the previous
discussions, we can derive the IR limit of the Anderson impurity model
by first integrating out both $D$ and $\tilde D$ fields, which amounts
to substituting
\beq
D^\dagger &=& -\frac{2}{U}\bar\theta\left(s\varphi^\dagger + \sum_{\vec k,\sigma} v_{\vec k} c_{\vec k,\sigma}^\dagger V_\sigma a^\dagger_\sigma\right)\\
\tilde D &=& -\frac{2}{U}\left(\tilde s\varphi + \sum_{\vec k,\sigma} v_{\vec k} c_{\vec k,\sigma}^\dagger V_\sigma a^\dagger_\sigma\right)\theta
\eeq
into $H$. We finally obtain
\beq
H_{IR}&=& H_{\rm Kondo} -\frac{2}{U}(s^2 \varphi^\dagger \varphi +\tilde s^2 \tilde \varphi^\dagger \tilde \varphi) - ( s\varphi^\dagger + \tilde s \tilde \varphi^\dagger )a_\uparrow a_\downarrow \nonumber\\ &-&\frac{2}{U}(s\varphi^\dagger + \tilde s \tilde \varphi)\sum_{\vec k, \sigma} v_{\vec k} a_\sigma V_\sigma c_{\vec k,\sigma} + h.c.
\eeq
where $H_{\rm Kondo}$ is the Kondo Hamiltonian
\beq
H_{\rm Kondo} &=& \sum_{\vec k,\sigma} n_{\vec k,\sigma} \varepsilon_{\vec k} -\frac{U}{2} (n_\uparrow + n_\downarrow)\nonumber\\
&& +\sum_{\vec k, \vec k'} \frac{4v_{\vec k} v_{\vec k'} }{U} \left[ (c_{\vec k}^\dagger \overrightarrow{\sigma} c_{\vec k'} )\cdot \vec S_{imp}\right.\nonumber\\
&&\left. - \frac{1}{2}c_{\vec k}^\dagger c_{\vec k'} (n_\uparrow+ n_\downarrow)  \right].
\eeq
Here, $\vec S_{imp}= a^\dagger_\sigma \overrightarrow
\sigma_{\sigma\sigma'} a_\sigma'$ is the spin operator of the impurity
site.
Thus the IR limit consists of the Kondo model in addition to the
coupling between the electron and the new bosonic degree of freedom.

\subsubsection{Half-filled Hubbard Model}

Next, we perform the same procedure for the Hubbard model at
half-filling by introducing two new fermionic fields $D$ and $\tilde D$
associated with the double occupancy and double holes, respectively. We
consider the generalised Lagrangian
\beq
\label{half-filling}
L&=&\int d^2\theta \left[\sum_i (D_i^\dagger \dot D_i+\dot{\tilde D}_i^\dagger{\tilde D}_i)\right.\nonumber\\
&&\left.-\frac{t}{2} \sum_{i,j,\sigma} g_{ij} \left(D_j^\dagger \theta c_{i,\sigma}V_\sigma c_{j,-\sigma} +\bar\theta c_{i,\sigma}V_\sigma c_{j,-\sigma} \tilde D_j +h.c. \right)\right.\nonumber\\
&&\left.+\frac{U}{2}\sum_j (D_j^\dagger D_j - \tilde D_j \tilde D_j^\dagger) +H_{con}
\right]
\eeq
with the constraint terms given by
\beq
H_{con}&=& s\bar\theta\sum_i \varphi_i^\dagger (D_i - \theta c_{i,\uparrow} c_{i,\downarrow}) + h.c. \nonumber\\
&+& \tilde s \bar\theta\sum_i \tilde\varphi_i^\dagger (\tilde D_i - \theta c_{i,\uparrow}^\dagger c_{i,\downarrow}^\dagger) + h.c.
\eeq
Here, $\varphi_i$ and $\tilde \varphi_i$ are two the bosonic field
with charge $2e$ and $-2e$ respectively. Similar to the previous
result, if we first integrate out both the bosonic fields
$\varphi_i$, $\tilde \varphi_i$ and then $D_i$, $\tilde D_i$, the
Hubbard model is obtained and the generalised theory Eq.
\ref{half-filling} yields the correct UV limit. However, a different
IR limit is obtained if we first integrate out $D_i$ and $\tilde
D_i$,
\beq\label{hfham}
H^{\rm hf}_{\rm IR} &=& -\frac{t^2}{2U}\sum_{i}  \left( b_i^\dagger b_i  + b_i b_i^\dagger \right) -\sum_i (\frac{2 s^2}{U}\varphi_i^\dagger \varphi_i +
\frac{2 \tilde s^2}{U}\tilde \varphi_i^\dagger \tilde\varphi_i)\nonumber\\
&&+ \frac{t}{U}\sum_{i}(s\varphi_i^\dagger + \tilde s \tilde \varphi_i)b_i + h.c. \nonumber\\
&&-\sum_i \ (s\varphi_i^\dagger - \tilde s \tilde \varphi_i) c_{i,\uparrow} c_{i,\downarrow} + h.c.
\eeq
This Hamiltonian is invariant under the transformation
$c_{i,\sigma}\rightarrow \exp(i \vec Q \cdot \vec R_i)
c_{i,\sigma}^\dagger$, $\varphi_i \leftrightarrow \tilde \varphi_i$ and
$s \leftrightarrow \tilde s$. This invariant reflects the symmetry
between the double occupancy and the double hole in the system at half
filling. In contrast to the doped case as in Eqs. (\ref{HIR}) and
(\ref{HIRE}), no $\cal M$ matrices appear in the IR theory at half
filling.  Consequently, we arrive at a closed form for the low-energy
theory at half-filling.  The $b^\dagger b + b b^\dagger$ terms include a
spin-spin interaction as well as a three-site hopping term. However, at
half-filling, the three-site hopping term vanishes. As a result,charge
dynamics appear solely from motion of the charge $2e$ boson. This state
of affairs obtains because at half-filling, charge dynamics persists
only on the energy scale $U$. Since it is the boson that encodes the
high-energy scale, it stands to reason that only the boson term mediates
charge transport. As we show in Appendix A, it is the $\varphi_i$ terms
that break the spurious local SU(2) symmetry\cite{lsu2} of the
Heisenberg model and reinstate the global SU(2) symmetry of the Hubbard
model.  The local SU(2) symmetry in the Heisenberg model arose entirely
from projection.  This local SU(2) symmetry was noticed quite some time
earlier\cite{lsu2} but its origin was never clarified. In fact, it is
straightforward to check that in the lower-Hubbard band at half-filling,
all perturbative terms in $t/U$ only mediate spin physics and hence
preserve the local SU(2) symmetry of the Heisenberg model. Our
derivation (see Appendix B) lays plain that as a result of the boson
terms, this symmetry is absent from the true low-energy theory of the
Hubbard model.  Note, in the Anderson impurity models, the bosonic terms
play a similar role to destroy the local SU(2) symmetry that appears as
a result of projection.

Second, the true low-energy model preserves the sum rules associated
with the original model.  An essential property\cite{dzy,zeros} of the
half-filled Hubbard model when $U\gg t$ is the presence of a surface in
momentum space (the Luttinger surface) where the single-particle Green
function vanishes. Since the Mott state at half-filling has a gap, the
non-trivial implication of the zero surface is that the real part of the
Green function,
\beq\label{green}
R_\sigma(\vec k,0)=-\int_{-\infty}^{-\Delta_-}d\epsilon'\frac{ A_\sigma(\vec
k,\epsilon')}{\epsilon'}-\int_{\Delta_+}^\infty d\epsilon'
\frac{A_\sigma(\vec k,\epsilon')}{\epsilon'}
\eeq
vanishes. Here $A_\sigma(k,\epsilon)$ is the single-particle
spectral function which we are assuming to have a gap of width
$2\Delta$ symmetrically located about the chemical potential at
$\epsilon=0$.  Because $A(\vec k,\epsilon)>0$ away from the gap, and
$\epsilon$ changes sign above and below the gap, Eq. (\ref{green})
can pass through zero.  For this state of affairs to obtain, the
piecesof the integral below and above the gap must be retained.
Projected models which throw away the UHB fail to recover the zero
surface. What Eq. (\ref{hfham}) makes clear is that all the
information regarding the surface of zeros is now encoded into the
bosonic fields $\varphi_i$ and $\bar\varphi_i$. On the Luttinger
surface, the self-energy diverges, representing a break-down of
perturbation theory. As the bosonic fields cannot be obtained from
perturbation theory, we conclude that it is the emergence of the
bosonic field that accounts for the breakdown of Luttinger's
theorem\cite{zeros} and ultimately the Mott insulating state.

\subsection{Electron Operator}

In each of the low energy theories, the operator which creates a single
electron represents a composite excitation. To determine its form, we
add to each of the starting Lagrangians a source term that generates the
canonical electron operator when the constraint is solved.  The
appropriate transformation that yields the canonical electron operator
in the UV, is
\beq
L\rightarrow L+\sum_{i,\sigma} J_{i,\sigma}\left[\bar\theta\theta(1-n_{i,-\sigma} ) c_{i,\sigma}^\dagger - V_\sigma D_i^\dagger \theta c_{i,-\sigma}\right] +
 h.c.\nonumber
\eeq
However, in the IR in which we only integrate over the heavy degree of
freedom, $D_i$, the electron creation operator
\beq\label{cop}
c^\dagger_{i,\sigma}&\rightarrow&(1-n_{i,-\sigma})c_{i,\sigma}^\dagger - V_\sigma \frac{t}{U} b_j^\dagger{\cal M}_{ji}^{-1} c_{i,-\sigma}\nonumber\\
&+& V_\sigma \frac{s}{U}\varphi_j^\dagger{\cal M}_{ji}^{-1}c_{i,-\sigma}.
\eeq
For electron doping, we apply the generalized particle-hole transformation to obtain
\beq\label{cop}
c^\dagger_{i,\sigma}&\rightarrow& n_{i,-\sigma}c_{i,\sigma}^\dagger - V_\sigma \frac{t}{U} c_{i,-\sigma} {\cal M}_{ij}^{-1} b_j^\dagger\nonumber\\
&+& V_\sigma \frac{s}{U} c_{i,-\sigma} {\cal M}_{ij}^{-1} \tilde\varphi_j.
\eeq
as the generator of electron excitations in the IR. For either
doping, the electron operator contains the standard term for motion
in the LHB, $(1-n_{i,-\sigma})c_{i,\sigma}^\dagger$
($n_{i,-\sigma}c_{i,\sigma})$ in the UHB for electron doping) with a
renormalization from spin fluctuations (second term) and a new
charge $e$ excitation, $\varphi_j^\dagger{\cal
M}_{ji}^{-1}c_{i,-\sigma}$. Consequently, we predict that an
electron at low energies is in a superposition of the standard LHB
state (modified with spin fluctuations) and a new composite charge
$e$ state described by $c_{i,-\sigma}{\cal
M}_{ij}^{-1}\varphi_j^\dagger$. It is the presence of these two
distinct excitations that preserves the dynamical (hopping dependent)
part of the spectral weight transfer across the Mott gap. As shown in companion paper\cite{ftmexp}, there
are also experimental ramifications for the composite structure of
the electron.

At half-filling, a similar trick can be applied to generate the electron
operator.  In this case,
\beq
L\rightarrow L+\sum_{i,\sigma} J_{i,\sigma}\left[V_\sigma D_i^\dagger c_{i,-\sigma}\theta + V_\sigma\bar\theta c_{i,-\sigma}\tilde D_i \right] +
 h.c.\nonumber
\eeq
is the correct transformation to generate the canonical electron
operator in the UV.  If we now integrate the partition function over
$D_i$ and $\tilde{D}_i$, we find that the electron creation operator at
half-filling
\beq\label{cop1}
c_{i,\sigma}^\dagger\rightarrow - V_\sigma\frac{t}{U}\left(c_{i,-\sigma}b_i^\dagger + b_i^\dagger c_{i,-\sigma}\right)
+V_\sigma\frac{2}{U}\left(s \varphi_i^\dagger + \tilde s \tilde\varphi_i\right) c_{i,-\sigma}\nonumber
\eeq
has two important differences with its counterpart for $n\ne 1$.  First,
it lacks the standard LHB and UHB components as the chemical potential
lies between both bands at half-filling.  Second, the propagator ${\cal
M}$ is absent. Nonetheless, the electron at half-filling still has two
components both above and below the chemical potential.  The
simplification that $c_{i,-\sigma}\varphi_i^\dagger$ (that is, the
${\cal M}$ matrix is absent) constitutes the new charge $e$ excitation
may make subsequent calculations of the strength of the binding between
the boson and a hole at least tractable within the framework of the
Bethe-Saltpeter equations.

\section{Final Remarks}

We have shown that a true low-energy theory of a doped Mott insulator
possesses degrees of freedom which do not have the quantum numbers of
the electron.  The degree of freedom is a local non-retarded charge $2e$
boson and hence stands in stark contrast to the charge $e$ boson in the
slave boson\cite{Slave1,Slave2,Slave3} approach in which a direct
integration of the high energy scale is not possible. Fundamental to
theory here is that the boson does not act in its own Fock space, in
contrast to other Fermi-Bose models\cite{tdlee,ranninger}. That is,
there are no free charge 2e boson states just as there are no free quark
states in confining theories. Rather, the charge 2e boson mediates new
electronic states by forming composite excitations.  As such the charge
2e degree of freedom is detectable\cite{letter,ftmexp} through the
substructure it provides in the electron excitation spectrum.  In
addition, the boson is not minimally coupled to the electromagnetic
gauge field.  It acquires dynamics and hence a gauge coupling through
high order terms in the ${\cal M}$ matrix, essentially $t^3/U^2$.  At
half-filling, the bosonic mode preserves the Luttinger surface on which
the self-energy diverges or equivalently, the single-particle Green
function vanishes.  Since the boson represents a non-perturbative
effect, it is not surprising that the Luttinger surface cannot occur
without it.  In a future publication\cite{ftmexp} we explore the role of
the boson in mediating the normal state properties of the cuprates as
well as the possibility that the Mott state is ultimately characterized
by charge neutral bound states mediated by the hidden charge $\pm2e$
boson.

\section{Appendix A: Lagrange Multipliers}
\newcommand{\vs}{\langle\sigma\rangle}

Here we offer some details about the mechanics of Lagrange multipliers.
Though this is standard stuff, we review it to avoid any confusion in
our derivation. To illustrate the method, we will begin with the
familiar example of the non-linear $\sigma$-model, that is a bosonic
theory with spherical target manifold. In Lorentzian signature, we
introduce the spherical constraint by writing the corresponding
functional $\delta$-function as an integral of a complex exponential,
with Lagrange multiplier $\sigma$:
\begin{widetext}
\beq
Z_L[J]&=&\int [d\phi^a d\sigma]\ e^{i\int d^dx\ \left\{ \frac12\phi^a \partial^2\phi^a
\right\}}e^{-\frac{i}{2}\int d^dx\ \sigma \left(\phi^a\phi^a-N/g\right)}
\eeq
\end{widetext}
which after Wick rotation becomes
\beq
Z_E[J]&=&\label{eq:fulleucl}\int [d\phi^a d\sigma]\ e^{-\int d^dx\ \left\{ \frac12\phi^a (-\partial^2+\sigma)\phi^a-\frac{N}{2g}\sigma
\right\}}\\
&=&\label{eq:phiint} \int [d\sigma]\ e^{-NS_{eff}(\sigma
)}
\eeq
with
\beq\label{eq:effact}
S_{eff}(\sigma
)=\frac12 Tr\ln (-\partial^2+\sigma)-\frac{1}{2g}\int_x\ \sigma
\eeq
In eq. (\ref{eq:phiint}), we have performed the $\phi^a$ functional
integrations. To proceed with the analysis of the model, we investigate
(\ref{eq:effact}) by expanding $\sigma$ around its vev $\vs$. It is
crucial though to appreciate that if we go back to eq.
(\ref{eq:fulleucl}), we see that in order for $\sigma$ to be a Lagrange
multiplier field (in the Euclidean formulation), the fluctuations in
$\sigma$ should be taken along the imaginary axis in field space. That
is, we write
\beq\label{lm}
\sigma(x)=\vs+\frac{i}{\sqrt{N}}\lambda(x).
\eeq
For uniform $\vs$, we then obtain
\begin{widetext}
\beq
Tr\ln (-\partial^2+\sigma) =
-Tr\ln \Delta_0 + \frac{i}{\sqrt{N}}\Delta_0(0)\int_x \lambda(x)
+\frac{1}{2N}\int_{x,y} \Delta_0(x,y)\lambda(y)\Delta_0(y,x)\lambda(x)
+\ldots
\eeq
\end{widetext}
where $\Delta_0^{-1} = -\partial^2+\vs$.
We thus find
\begin{widetext}
\beq
S_{eff}=-\frac{\vs}{2g}V_d-\frac12 Tr\ln\Delta_0
-\frac{i}{2g\sqrt{N}}\int_x\lambda(x)\left(1-g\Delta_0(0)
\right)
+\frac{1}{4N}\int_{x,y} \Delta_0(x,y)\lambda(y)\Delta_0(y,x)\lambda(x)
+\ldots
\eeq
\end{widetext}
Thus we see that there is a stable saddle point giving the familiar gap equation
\beq
\frac{1}{g} = \int \frac{d^dp}{(2\pi)^d}\ \frac{1}{p^2+\vs}
\eeq

Now, in the theory considered in this paper, we have a similar situation.
In the Lorentzian signature, we have
\beq\label{Z}
Z=\int [{\cal D}c\ {\cal D}c^\dagger\ {\cal D}D\ {\cal D}D^\dagger\ {\cal D}\varphi\ {\cal D}\varphi^\dagger]\exp^{i\int d^dx (L_0+H_{\rm con}) },\nonumber\\
\eeq
where
\beq
L_0&=&\int d^2\theta\left[\bar{\theta}\theta\sum_{i,\sigma}(1- n_{i,-\sigma}) c^\dagger_{i,\sigma}\dot c_{i,\sigma} +\sum_i D_i^\dagger\dot D_i\right.\nonumber\\
&&+U\sum_j D^\dagger_jD_j- t\sum_{i,j,\sigma}g_{ij}\left[ C_{ij,\sigma}c^\dagger_{i,\sigma}c_{j,\sigma}
+D_i^\dagger c^\dagger_{j,\sigma}c_{i,\sigma}D_j\right.\nonumber\\
&&+\left.\left.(D_j^\dagger \theta c_{i,\sigma}V_\sigma c_{j,-\sigma}+h.c.)\right]\right].
\eeq
Here again, we have written the functional $\delta-$function constraint
as the integral of a complex exponential.  In this case, however,
$\varphi_i$ the Lagrange multiplier is a complex field such that
$\varphi_i^\ast=\Re\varphi_i-i\Im\varphi_i$.  After Wick rotation, we
obtain the path integral in Euclidean signature,
\beq
 Z=\int [{\cal D}c\ {\cal D}c^\dagger\ {\cal D}D\ {\cal D}D^\dagger\ {\cal D}\varphi\ {\cal D}\varphi^\dagger]\exp^{-\int d^dx (L_0+H_{\rm con}) }.\nonumber\\
\eeq
In either the Lorentz or Euclidean signatures, $\varphi_i$ is a Lagrange
multiplier.  That is, the integral over $\varphi_i$ must still yield a
functional $\delta-$function even though the $i$ is absent.  This
requirement dictates that we must integrate the fluctuations of both the
real and imaginary parts of $\varphi_i$ along the imaginary axis as in
Eq. (\ref{lm}).  The result is a stable Gaussian integral (if $D$ is
integrated first and then $\varphi_i$) which can be evaluated using Eq.
(\ref{iden}).  Of course in the reverse order, the $\varphi_i$ integrals
simply yield $\delta-$ functions.  In both cases, one ends up with the
$+U$ Hubbard model.

\section{Appendix B: Absence of Local SU(2) Symmetry}
Here we consider the possibility of trivially extending our model at
half-filling to a locally SU(2) invariant theory.
%
Let us organize the electron operators in the form
\beq
\Psi = \left(\begin{array}{cc}
                            c_\ua & c_\da \\
                            c^\dagger_\da & -c^\dagger_\ua
                            \end{array} \right)
\eeq
where the spatial index is suppressed.  A local SU(2) transformation
acts via left multiplication by an SU(2) matrix $h$ so that $\Psi
\rightarrow h \Psi$.  Let us write $h$ as
\beq
h = \left(\begin{array}{cc}
                        \alpha & \beta \\
                        -\beta^\ast & \alpha^\ast
                        \end{array} \right)
\eeq
where $|\alpha|^2 + |\beta|^2 = 1$.  The electron bilinear $c_\ua c_\da$
is a member of the triplet of the local $SU(2)$. We will now show that
the correct ``middle" term in the electron triplet is $(n_\ua + n_\da -
1)/\sqrt{2}$.  Transforming the prospective triplet we find
\begin{widetext}
\beq
\left(\begin{array}{c}
                c_\da c_\ua \\
                \frac{n_\ua + n_\da - 1}{\sqrt{2}} \\
                c^\dagger_\ua c^\dagger_\da
                \end{array} \right)
                \rightarrow
\left(\begin{array}{ccc}
                \alpha^2 & \sqrt{2} \alpha \beta^\ast & -\beta^{\ast 2} \\
                -\sqrt{2} \alpha \beta & |\alpha|^2 - |\beta|^2 & -\sqrt{2} \alpha^\ast \beta^\ast \\
                -\beta^2 & \sqrt{2} \alpha^\ast \beta & \alpha^{\ast 2}
                \end{array} \right)
\left(\begin{array}{c}
                c_\da c_\ua \\
                \frac{n_\ua + n_\da - 1}{\sqrt{2}} \\
                c^\dagger_\ua c^\dagger_\da
                \end{array} \right)
\eeq
\end{widetext}
By inspection it is clear that this matrix is unitary since $h$ is
unitary, and also that its determinant is $(|\alpha|^2 + |\beta|^2)^3 =
1$.  Thus, we have an SU(2) transformation and we conclude that the
electron triplet is
\beq
W = \left(\begin{array}{c}
              c_\da c_\ua \\
                \frac{n_\ua + n_\da - 1}{\sqrt{2}} \\
                c^\dagger_\ua c^\dagger_\da
                \end{array} \right)
\eeq
%

Thus, to make our theory local $SU(2)$ invariant, we must also take $D$
and $\varphi$ to reside in triplets as well. Thus, we posit a charge
zero bosonic field $\varphi_0$ which we collect into a triplet $\Phi =
(\varphi,\varphi_0,\tilde{\varphi})^T$.  $SU(2)$ invariance would
necessitate an additional constraint term of the form
\beq
H_{con} \rightarrow H_{con} + s \bar{\theta} \sum_i \varphi^\dagger_{0 i} (D_{0i} - \theta (n_\ua + n_\da - 1))
\eeq
%

Just as in the cases of the earlier boson fields, the new field
$\varphi_0$ enters the theory as a Lagrange multiplier corresponding to
a dynamical field $D_0$, which will be integrated over.  Now that we
have determined the form of the electron triplet we can write the
generalized Lagrangian, now including the $\varphi_0$ terms, in the form
\begin{widetext}
\beq\label{sutwolag}
L&=&\int d^2\theta \left[\sum_i (aD_i^\dagger \dot D_i+b\dot{\tilde D}_i^\dagger {\tilde D}_i+c D_{0i}^\dagger \dot D_{0i})+\frac{1}{2}U\sum_j (\alpha D_j^\dagger D_j - \beta\tilde D^\dagger_j \tilde D_j+\frac{\delta}{2}D_{0j}^\dagger D_{0j})\right.\nonumber\\
\nonumber\\
&&\left.-t \sum_{i,j,\sigma} g_{ij}\frac12 \left(D_j^\dagger \theta c_{i,\sigma}V_\sigma c_{j,-\sigma} -\bar\theta c_{i,\sigma}V_\sigma c_{j,-\sigma} \tilde D_j+\bar\theta D_{0i}c^\dagger_{i,\sigma}c_{j,\sigma} +h.c. \right)+H_{\rm con}
\right]
\eeq
where the constraint is given by
\beq
H_{con}&=& s\bar\theta\sum_i \varphi_i^\dagger (D_i - \theta c_{i,\uparrow} c_{i,\downarrow}) + h.c. + \tilde s \bar\theta\sum_i \tilde\varphi_i^\dagger (\tilde D_i -\theta c_{i,\uparrow}^\dagger c_{i,\downarrow}^\dagger) + h.c.+s\bar\theta\sum_i\varphi_{0i}^\dagger (D_{0i}-\theta(n_{i,\uparrow}+n_{i,\downarrow}-1))
\eeq
The undetermined coefficients may be fixed by the condition that the
theory reduces to the Hubbard model when the constraints are solved. 
Integrating over the $\varphi_i$ and then $D$ fields, yields
\begin{eqnarray}
L & \rightarrow &\sum_i \left[a n_{i -\sigma} c^\dagger_{i \sigma} \dot{c}_{i \sigma} - b(1-n_{i -\sigma}) c^\dagger_{i \sigma} \dot c_{i \sigma}  \right]
+ 
\frac{U}{2} \sum_i \left[(\beta - \frac{\delta}{2}) (n_{i \ua} + n_{i \da} - 1) + (\alpha - \beta + \delta) n_{i \ua} n_{i \da} \right]
\end{eqnarray}
up to total time derivatives.
%
In order for this to yield the Hubbard model we must therefore have
$\beta = 1 - \alpha$ and $\delta = 2(1-\alpha)$.  We note that
$\alpha=\beta=2\delta=1/2$ solves these, and is in fact the $SU(2)$
symmetric point. Similarly, we find $a=-b=1$. The coefficient $c$ is
unconstrained because $D_0$ has trivial dynamics.
We have found then that the theory at half filling may be extended to an
$SU(2)$ invariant theory. This $SU(2)$ acts only globally however; this
may be plainly seen by examining the interaction terms on the second
line of eq. (\ref{sutwolag}). In order to make these terms local $SU(2)$
invariant, an explicit $SU(2)$ gauge field (a Wilson line) would have to
be introduced.
We conclude, then, that the local SU(2) symmetry of the Heisenberg model
is broken by the presence of the bosonic degrees of freedom in our
model.  That a local $SU(2)$ symmetric version of the theory cannot be
constructed is not surprising as the Hubbard model lacks this symmetry
-- in that case it is broken by hopping terms (which again could only be
made invariant by the introduction of explicit Wilson lines).
%
In the case of the Heisenberg model, the local $SU(2)$ symmetry appears
strictly because of projection.  Hence,  the exact low-energy theory
constructed without using projection should not possess symmetries not
found in the Hubbard model.
\end{widetext}

\acknowledgements We thank D. Haldane for helpful comments regarding the
range of validity of the $t-J$ model, E. Fradkin for pointing out the
SU(2) references, and the NSF DMR-0605769 for partial support.


\begin{thebibliography}{99}
\bibitem{charge}In this expression for $Q$, $\varphi^\dagger$ should be interpreted as a creation operator with the appropriate charge.
\bibitem{complex}A complex HS transformation of the type in Eq. (\ref{iden}) is closely related to the exponentiation of a $\delta$-functional constraint.
\end{thebibliography}
\end{document}